\documentclass[pra,aps,epsfig,psfig,multicols,showpacs,tightenlines]{revtex4}

\usepackage{graphics,bm}
\usepackage{graphicx}

\newcommand{\beq}{\begin{equation}}
\newcommand{\eeq}{\end{equation}}
\newcommand{\bqa}{\begin{eqnarray}}
\newcommand{\eqa}{\end{eqnarray}}
\def\gsim{\mathrel {\vcenter {\baselineskip 0pt \kern 0pt
\hbox{$>$} \kern 0pt \hbox{$\sim$} }}}

\protect

\begin{document}
\title{Soliton localization in Bose-Einstein condensates with time-dependent
harmonic potential and scattering length}
\author{Usama Al Khawaja}
\affiliation{ \it Physics Department, United Arab Emirates University, P.O. Box 17551, Al-Ain, United Arab
Emirates.}

\date{\today}

\begin{abstract}
We derive exact solitonic solutions of a class of Gross-Pitaevskii
equations with time-dependent harmonic trapping potential and
interatomic interaction. We find families of exact single-solitonic,
multi-solitonic, and solitary wave solutions. We show that, with the
special case of an oscillating trapping potential and interatomic
interaction, a soliton can be localized indefinitely at an arbitrary
position. The localization is shown to be experimentally possible
for sufficiently long time even with only an oscillating trapping
potential and a constant interatomic interaction.
\end{abstract}


\maketitle

\section{ Introduction}

The experimental realization of solitons in Bose-Einstein
condensates \cite{burger,denschlag,anderson,randy,schreck,eirman}
has stimulated intense interest in their properties
\cite{perez1,busch,linear1,cast,linear2,kasa,sala0,abdu,sala}. The
inhomogeneity provided by the trapping potential has renewed the old
\cite{zab,has0,chen,konotop1,franco} and more recent \cite{serk}
interest in the different aspects of one- and multi-solitons
dynamics in inhomogeneous potentials.

In the experiments of Refs.~\cite{randy,schreck}, stable bright
solitons were created and set in a particle-like center-of-mass
motion. The wave nature of solitons was revealed when two adjacent
solitons repelled each other as a result of their phase difference
\cite{linear1}. On the other hand, it is established that bright
solitons collapse when the number of atoms exceeds a certain limit
\cite{book}. This results from the attractive Hartree energy
overcoming the repulsive kinetic energy pressure. One of the methods
proposed to stabilize the soliton against collapsing is to rapidly
oscillate the interatomic interaction or the trapping potential
\cite{vibrating}. Obviously, the inhomogeneity imposed by the
trapping potential plays an important role on the dynamics and
stability of solitons.

The evolution of solitons is approximately described by the
inhomogeneous nonlinear Schr$\ddot{\rm o}$dinger equation known as
the Gross-Pitaeviskii equation \cite{gpref1,gpref2}. The
approximation stems from the fact that the Gross-Pitaevskii equation
is a mean-field approximation of the exact $N$-particle
Schr$\ddot{\rm o}$dinger equation. While in some cases the soliton
dynamics obtained by these two equations disagree~\cite{castan}, the
Gross-Pitaevskii equation often gives accurate results. Theoretical
studies performed to account for the observed behavior of solitons
were conducted by solving the Gross-Pitaevskii equation with
numerical, perturbative, or variational methods. Much less effort
was devoted to finding exact solutions of this equation
\cite{chen,wu,ieee,jun,liang,lu,raj,usamapre,ramesh}. In addition to
providing rigorous insight, exact solutions acquire valuable
importance when problems such as soliton-soliton collisions and
soliton interaction with potentials are addressed. In such cases,
formulae for the force between solitons or soliton's effective mass
can be derived \cite{gordon}. In addition, obtaining such exact
solutions allows for testing the validity of the Gross-Pitaevskii
equation at high soliton densities and obtaining the long-time
evolution of the soliton where numerical techniques aught to break
down.

Here, we further explore exact solitonic solutions of the
Gross-Pitaevskii equation. Specifically, the goal of this paper is
two-fold. First, we investigate the existence and properties of
solitonic solutions in the presence of time-dependent trapping
potential and interatomic interaction. Secondly, we focus on the
effect of an oscillating trapping potential and interatomic
interaction on the center-of-mass motion of the soliton. We consider
here only the case of attractive interatomic interactions which
allows for bright solitons.

The first goal is achieved by employing the Darboux transformation
method \cite{salle} to derive families of exact solitonic solutions
of a class of Gross-Pitaevskii equations. It should be noted that
the main solution we derive here (Eq.~(\ref{psi_exact})), which
corresponds to a harmonic expulsive potential, reproduces a special
case of the general solution found by Serkin {\it et al.}
corresponding to a combination of a harmonic and linear
potentials~\cite{serk}. Hence, the significance of the first goal is
mainly presenting a systematic method of generating exact solutions.

For the second goal, we considered the special case of oscillating
strengths of the harmonic trapping potential and interatomic
interaction. Interestingly enough, it turns out that such
oscillations not only stabilize the soliton against shrinking, but
also make it possible to localize it at an arbitrary position. This
is the main result of this paper. The possibility of localizing the
soliton is then discussed from an experimental point of view. To
that end, we considered the nonintegrable, though experimentally
simpler, case of only an oscillating trapping potential and constant
interatomic interaction. Here too, the soliton can be localized,
though not indefinitely as before. The long-time evolution shows
that the soliton continues to be trapped but will be oscillating
around the minimum of the harmonic potential. For a typical
experimental setup, we show that the soliton can be localized around
its initial position for a time period of the order of, or even
larger than, the lifetime of the soliton. The soliton localization
suggests a management mechanism for the soliton position and speed
that may have applications in various situations such as
soliton-soliton collisions and soliton interaction with potentials.

The rest of the paper is organized as follows. In the next section,
we present the general form of the Gross-Pitaevskii equation to be
solved. In section~\ref{darboux}, we use the Darboux transformation
method to derive the new solitonic solutions. We then discuss their
properties, dynamics, and localization. In section~\ref{secnum}, we
discuss the experimental feasibility of realizing soliton
localization. We end in section~\ref{conc} with a summary of our
main results and conclusions.

\section{The Gross-Pitaevskii equation}
\label{gpsec}

The Gross-Pitaevskii equation describing a Bose-Einstein condensate
trapped by an axially-symmetric harmonic potential with attractive
interatomic interactions is given by
\begin{equation}
i\hbar{\partial\over\partial t}\psi({\bf
r},t)=\left[-{\hbar^2\over2m}\nabla_{\bf
r}^2+{1\over2}m\,\left(\omega_{x}^2\,x^2+\omega_{\perp}^2(y^2+z^2)\right)
-{4\,\pi\,a_s\,\hbar^2\over m}|\psi({\bf r},t)|^2\right]\psi({\bf
r},t) \label{gp3d},
\end{equation}
where $a_s$ is the absolute value of the $s$-wave scattering length,
and $\omega_x$ and $\omega_\perp$ are the characteristic frequencies
of the harmonic trapping potential in the axial and radial
directions, respectively.

When the confinement of the Bose-Einstein condensate is much larger
in the $y$ and $z$ directions compared to the confinement in the $x$
direction, the system can be considered effectively one-dimensional
along the $x$ direction. The three-dimensional Gross-Pitaevskii
equation can then be {\it integrated} over the $y$ and $z$
directions to result in a one-dimensional Gross-Pitaevskii equation
\cite{perez1,carr}
\begin{equation}
i\hbar{\partial\over\partial
t}\psi(x,t)=\left[-{\hbar^2\over2m}{\partial^2\over\partial
x^2}+{1\over2}m\,\omega_x^2\,
x^2-2\lambda\,a_s|\psi(x,t)|^2\right]\psi(x,t) \label{gpscaled0},
\end{equation}
where $\lambda=\omega_\perp/\omega_x$. Scaling length to
$a_{x}=\sqrt{\hbar/m\omega_x}$, time to $1/\omega_x$, and
$\psi(x,t)$ to $1/\sqrt{2\lambda\,a_x}$, the Gross-Pitaevskii
equation takes the dimensionless form
\begin{equation}
i{\partial\over\partial
t}\psi(x,t)=\left[-{1\over2}{\partial^2\over\partial
x^2}+{1\over2}\,p(t)\, x^2-a\,q(t)|\psi(x,t)|^2\right]\psi(x,t)
\label{gpscaled},
\end{equation}
where $a=a_s/a_x$ is the scaled scattering length. The dimensionless
general functions $p(t)$ and $q(t)$ are introduced to account for
the time-dependencies of the strengths of the trapping potential and
the interatomic interaction.

In the subsequent section, it is shown that this equation is
integrable only if $p(t)$ and $q(t)$ are parametrically related as
follows: $p(t)={\ddot \gamma}(t)-{\dot\gamma}(t)^2$ and
$q(t)=\exp{(\gamma(t))}$, where $\gamma(t)$ is an arbitrary real
function (see Eq.~(\ref{gp2})). We find exact solitonic solutions of
Eq.~(\ref{gpscaled}) below in terms of the function $\gamma(t)$. A
host of rich and interesting physical systems are described by such
a class of solutions. In particular, an oscillating form of
$\gamma(t)$ will be considered in this paper.

\section{Darboux Transformation and the Exact Solutions}
\label{darboux}
\subsection{Darboux Transformation}
\label{subsecdarboux}

Applying the Darboux transformation method on nonlinear partial
differential equations requires finding a linear system of equations
for an {\it auxiliary} field ${\bf \Psi}(x,t)$. The linear system is
usually written in a compact form in terms of a pair of matrices as
follows: ${\bf \Psi}_x={\bf U}\cdot{\bf \Psi}$ and ${\bf
\Psi}_t={\bf V}\cdot{\bf \Psi}$. The matrices ${\bf U}$ and ${\bf
V}$, known as the Lax pair, are functionals of the solution of the
differential equation. The {\it consistency condition} of the linear
system ${\bf \Psi}_{xt}={\bf \Psi}_{tx}$ is required to be
equivalent to the partial differential equation under consideration.
Applying the Darboux transformation, as defined below, on ${\bf
\Psi}$ transforms it into another field ${\bf \Psi}[1]$. For the
transformed field ${\bf \Psi}[1]$ to be a solution of the linear
system, the Lax pair must also be transformed in a certain manner.
The transformed Lax pair will be a functional of a new solution of
the same differential equation.

Practically, this is performed as follows. First, we find the Lax
pair and an exact solution of the differential equation, known as
the {\it seed} solution. Fortunately, the trivial solution can be
used as a seed, leading to nontrivial solutions. Using the Lax pair
and the seed solution, the linear system is then solved and the
components of ${\bf \Psi}$ are found. The new solution is expressed
in terms of these components and the seed solution. The following
detailed derivation of the new solution clarifies this procedure
further.

Using our Lax Pair search method \cite{usama_darboux}, we find the
following linear system which corresponds to the class of
Gross-Pitaevskii equations we are interested in:
\begin{equation}
{\bf \Psi}_x=\zeta{\bf J\cdot\Psi\cdot\Lambda}+{\bf U\cdot \Psi}
\label{psi_x},
\end{equation}
\begin{equation}
{\bf \Psi}_t=i\zeta^2{\bf J\cdot\Psi\cdot
\Lambda\cdot\Lambda}+\zeta\,\left(i{\bf U}+ x\, \gamma(t)\,{\bf
J}\right) {\bf\cdot\Psi\cdot\Lambda}+{\bf V\cdot \Psi}
\label{psi_t},
\end{equation}
where,\\\\
\begin{math}
{\bf \Psi}(x,t)=\left(\begin{array}{cc}
\psi_1(x,t)&\psi_2(x,t)\\
\phi_1(x,t)&\phi_2(x,t)
\end{array}\right)
\end{math},
\hspace{0.25cm}
\begin{math}
{\bf J}=\left(\begin{array}{cc}
1&0\\
0&-1
\end{array}\right)
\end{math},
\hspace{0.25cm}
\begin{math}
\\\\\\
{\bf \Lambda}=\left(\begin{array}{cc}
\lambda_1&0\\
0&\lambda_2
\end{array}\right)
\end{math},
\hspace{0.25cm}
\begin{math}
{\bf U}=\left(\begin{array}{cc}
0&\sqrt{a}Q(x,t)\\
-\sqrt{a}Q^*(x,t)&0
\end{array}\right)
\end{math},
\\\\\\
\begin{math}
{\bf V}=\left(\begin{array}{cc}ia|Q(x,t)|^2/2&\sqrt{a}\lambda x
{\dot\gamma}Q(x,t)
+i\sqrt{a}Q_x(x,t)/2\\
-\sqrt{a}\lambda x
{\dot\gamma}Q^*(x,t)+i\sqrt{a}Q^*_x(x,t)/2&-ia|Q(x,t)|^2/2
\end{array}\right)
\end{math},\\\\\\
$\zeta(t)=\exp{\left(\gamma(t)\right)}$, and $\lambda_1$ and
$\lambda_2$ are arbitrary constants.  For convenience, we presented
the matrices in terms of the function $Q(x,t)$ which is related to
the wave function as follows
$Q(x,t)=\psi(x,t)e^{(\gamma(t)+i{\dot\gamma}(t)x^2)/2}$.

It should be emphasized that while applying the Darboux
transformation is almost straightforward, finding a linear system
that corresponds to the differential equation at hand is certainly
not a trivial matter. Usually, this is found by trial and error, or
by starting from a certain linear system and then finding the
differential equation it corresponds to. In
Ref.~\cite{usama_darboux}, we have introduced a systematic approach
to find the linear system which we describe here briefly. The
partial derivatives of the auxiliary field, ${\bf \Psi}_x$ and ${\bf
\Psi}_t$, are expanded in powers of $\bf\Lambda$ with unknown matrix
coefficients. The expansions are terminated at the first order for
${\bf \Psi}_x$ and the second order for ${\bf \Psi}_t$ since this
will be sufficient to generate the class of Gross-Pitaevskii
equations under consideration. The higher order matrix coefficients
turn out to be essentially determined  by the zeroth order matrix
coefficients $\bf U$ and $\bf V$. To find the matrices $\bf U$ and
$\bf V$, we expand them in powers of the wavefunction $\psi(x,t)$,
its complex conjugate, and their partial derivatives. The
coefficients of the expansions are unknown functions of $x$ and $t$.
Substituting these expansions in the consistency condition
(Eq.~(\ref{consis}) below) we find a set of equations for the
unknown function coefficients. Finally, by solving these equations
the Lax pair and consequently the linear system will be determined.
The linear system found here is a generalization to that of
Zakharov-Shabat for homogeneous Gross-Pitaevskii equation \cite{zs}.

For ${\bf\Psi}$ to be a solution of both Eqs.~(\ref{psi_x}) and
(\ref{psi_t}), the consistency condition
${\bf\Psi}_{xt}={\bf\Psi}_{tx}$ must be satisfied. This condition
leads to the following relation between the matrices $\bf U$ and
$\bf V$
\begin{equation}
{\bf U}_t-{\bf V}_x+[{\bf U},{\bf V}]=0 \label{consis},
\end{equation}
where $[{\bf U},{\bf V}]$ is the commutator of ${\bf U}$ and ${\bf V}$. Substituting the above expressions for
$\bf U$ and $\bf V$ in the last equation, we obtain the Gross-Pitaevskii equation
\begin{equation}
i{\partial\over\partial
t}\psi(x,t)=\left[-{1\over2}{\partial^2\over\partial x^2}+{1\over2}
\left({\ddot \gamma}(t)-{\dot\gamma}(t)^2\right)x^2\psi(x,t)-a\,
e^{{\gamma(t)}}|\psi(x,t)|^2\right]\psi(x,t) \label{gp2},
\end{equation}
and its complex conjugate. This equation shows that the functions
$p(t)$ and $q(t)$ are parametrically related to each other through
the general function $\gamma(t)$. In Ref.~\cite{serk}, Serkin {\it
et al.} solve a nonautonomous Gross-Pitaevskii equation that is
similar to Eq.~(\ref{gp2}) but with additional time-dependent
dispersion and linear potential. Similar to our conclusion
\cite{usama_darboux}, the authors of this reference and
Ref.~\cite{ramesh} found that relations between the coefficients
must be obeyed for the the model to be integrable. We focus here on
the specific case of only a harmonic trapping potential and constant
dispersion. The linear system of 8 equations, Eqs.~(\ref{psi_x}) and
(\ref{psi_t}), read explicitly
\begin{equation}
\psi _{1x}-\phi _1 \sqrt{a}\, e^{\frac{1}{2} \left(i \dot{\gamma }
x^2+\gamma \right)} \psi _0-\sqrt{2} \lambda _1 \psi _1 e^{\gamma }
=0\label{l1},
\end{equation}

\begin{equation}
\psi _{2x}-\phi _2 \sqrt{a}\, e^{\frac{1}{2} \left(i \dot{\gamma }
x^2+\gamma \right)} \psi _0-\sqrt{2} \lambda _2 \psi _2 e^{\gamma }
=0
  \label{l2},
\end{equation}

\begin{equation}
\phi _{1x}+\sqrt{2} \lambda _1\, e^{\gamma } \phi _1+\psi _0^* \psi
_1 \sqrt{a}\, e^{\frac{1}{2} \left(\gamma -i x^2 \dot{\gamma
}\right)}=0
    \label{l3},
\end{equation}

\begin{equation}
\phi _{2x}+\sqrt{2} \lambda _2\, e^{\gamma } \phi _2+\psi _0^* \psi
_2 \sqrt{a}\, e^{\frac{1}{2} \left(\gamma -i x^2 \dot{\gamma
}\right)} =0
    \label{l4},
\end{equation}

\begin{eqnarray}
 \psi _{1t}-i \psi _1 e^{\gamma } \left(2 e^{\gamma } \lambda
_1^2- i \sqrt{2} x \dot{\gamma } \lambda _1+{1\over2}\left|\psi
   _0\left|^2\right.\right. a\right)-{1\over2}\phi _1 \sqrt{a}\, e^{\frac{1}{2} \left(i \dot{\gamma } x^2+\gamma \right)} \left(i \psi _{0x}+\psi _0
   \left(2 i \sqrt{2} e^{\gamma } \lambda _1+x \dot{\gamma }\right)\right)=0
     \label{l5},
\end{eqnarray}

\begin{eqnarray}
 \psi _{2t}-i \psi _2 e^{\gamma } \left(2 e^{\gamma } \lambda
_2^2- i \sqrt{2} x \dot{\gamma } \lambda _2+{1\over2}\left|\psi
   _0\left|^2\right.\right. a\right)-{1\over2}\phi _2 \sqrt{a}\, e^{\frac{1}{2} \left(i \dot{\gamma } x^2+\gamma \right)} \left(i \psi _{0x}+\psi _0
   \left(2 i \sqrt{2} e^{\gamma } \lambda _2+x \dot{\gamma }\right)\right)=0
     \label{l6},
\end{eqnarray}

\begin{eqnarray}
 \phi _{1t}+{1\over2}\psi _1 \sqrt{a}\, e^{\frac{1}{2} \left(\gamma -i x^2
\dot{\gamma }\right)} \left(\psi _0^* \left(2 i \sqrt{2} e^{\gamma }
   \lambda _1+x \dot{\gamma }\right)-i \psi _{0x}^*\right)+{1\over2}\phi _1 e^{\gamma } \left(i \left|\psi _0\left|^2\right.\right. a+2 \lambda _1
   \left(2 i e^{\gamma } \lambda _1+\sqrt{2} x \dot{\gamma }\right)\right)=0
     \label{l7},
\end{eqnarray}

\begin{eqnarray}
 \phi _{2t}+{1\over2}\psi _2 \sqrt{a}\, e^{\frac{1}{2} \left(\gamma -i x^2
\dot{\gamma }\right)} \left(\psi _0^* \left(2 i \sqrt{2} e^{\gamma }
   \lambda _2+x \dot{\gamma }\right)-i \psi _{0x}^*\right)+{1\over2}\phi _2 e^{\gamma } \left(i \left|\psi _0\left|^2\right.\right. a+2 \lambda _2
   \left(2 i e^{\gamma } \lambda _2+\sqrt{2} x \dot{\gamma }\right)\right)=0
     \label{l8},
\end{eqnarray}
where $\psi_0(x,t)$ is an exact seed solution of Eq.~(\ref{gp2}).
These equations reduce to an equivalent system of 4 equations with
nontrivial solutions by making the following substitutions:
$\lambda_1=-\lambda_2^*$, $\psi_2=\phi_1^*$, and $\phi_2=-\psi_1^*$.
Using the trivial solution, $\psi_0(x,t)=0$, as a seed, the linear
system will have the solution
\begin{eqnarray}
\psi_1(x,t)&=&{c_1} e^{2 i \lambda_1^2\int e^{2   \gamma(t)} \,
dt+e^{
   \gamma(t)}{\lambda_1} x }\label{ls1},
\\
\phi_2(x,t)&=& {c_2} e^{-2 i \lambda_1^2\int e^{2   \gamma(t)} \,
dt-e^{
    \gamma(t)} {\lambda_1}x }
\label{ls2},
\end{eqnarray}
where $c_1$ and $c_2$ are real arbitrary constants of integration.

Consider the following version of the Darboux
transformation~\cite{salle}
\begin{equation}
{\bf\Psi}[1]={\bf\Psi}\cdot{\bf\Lambda}-\sigma{\bf\Psi} \label{dt},
\end{equation}
where ${\bf\Psi}[1]$ is the transformed field,
$\sigma={{\bf\Psi_0}}\cdot{\bf\Lambda}\cdot{\bf\Psi_0}^{-1}$. Here
$\bf\Psi_0$ is a known (seed) solution of the linear system
Eqs.~(\ref{l1}-\ref{l8}). For the transformed field ${\bf\Psi}[1]$
to be a solution of the linear system, the matrix $\bf U$ for
instance must be transformed as \cite{note}
\begin{equation}
{\bf U}[1]=\sigma\cdot{\bf
U}\cdot\sigma^{-1}+\sigma_x\cdot\sigma^{-1}\label{u0p},
\end{equation}
where $\sigma^{-1}$ is the inverse of $\sigma$. This equation gives
the new solution in terms of the seed solutions of the
Gross-Pitaevskii equation, $\psi_0(x,t)$, and the linear system,
${\bf\Psi_0}$, which reads
\begin{equation}
\psi(x,t)=\psi_0(x,t)+{2\over a}(\lambda_1+\lambda_1^*)e^{-i
{\dot\gamma}(t)
x^2/4+\gamma(t)/2}{\phi_1\psi_1^*\over|\phi_1|^2+|\psi_1|^2}
\label{psinew}.
\end{equation}

\subsection{Exact single-solitonic solution}
\label{subsecexact} Using $\psi_0(x,t)=0$ and Eqs.~(\ref{ls1}) and
({\ref{ls2}}), the new solution takes the form
\begin{equation}
\psi(x,t)=\frac{4\sqrt{2}\,{\lambda_{1r}}\,{c_1}\,{{{c_2}}^*}}{{\sqrt{{a}}}}\frac{e^{
\gamma (t)/2+i\theta(t)}
}{|c_1|^2e^{\beta(t)x-x_0(t)}+|c_2|^2e^{-(\beta(t)x-x_0(t))}
}\label{psinew2},
\end{equation}
where $\theta(t)=-4\,\left( {{\lambda_{1i}}}^2 - {{\lambda_{1r}}}^2
\right)\,\int e^{2 \,\gamma (t)}\,dt {+x\,\left( 2\,e^{
\,\gamma(t)}\,{\lambda_{1i}} - x \,{\dot\gamma}(t)/\sqrt{8} \right)
}$, $\beta(t)=2\sqrt{2}{\lambda_{1r}}\,e^{ \gamma(t)}$, and
$x_0(t)=8\,{\lambda_{1i}}\,{\lambda_{1r}}\,\int e^{2 \,\gamma
              (t)}\,dt$.
Here $c_1$, $c_2$, $\lambda_1$, and $\lambda_2$ are arbitrary
constants. The subscripts $r$ and $i$ denote real and imaginary
parts, respectively. Substituting $c_1=\exp{(\delta_1)}$,
$c_2=\exp{(\delta_2)}$, where $\delta_1$ and $\delta_2$ are
arbitrary constants, completing the square in the phase factor, and
normalizing $\psi(z,t)$ to $N$, this solution can be recast in the
following more appealing form
\begin{equation}
\psi(x,t)={\sqrt{N}\sqrt{N\,a}\over2}\,e^{\gamma(t)/2+i\phi(x,t)}{\rm
sech}{\left({N\,a\over2}\,e^{\gamma(t)}(x-x_{\rm cm}(t))\right)}
\label{psi_exact},
\end{equation}
where
\begin{equation}
\phi(x,t)=\phi_0(t) +{\dot x_{\rm cm}}(t)\,(x-x_{\rm
cm}(t))-{1\over2}{\dot\gamma}(t)(x-x_{\rm cm}(t))^2 \label{pha},
\end{equation}
\begin{equation}
x_{\rm
cm}(t)=\left(x_0e^{2\gamma(0)}+\left(v_0+x_0\,{\dot\gamma}(0)\right)
g(t)\right)e^{-\gamma(t)-\gamma(0)} \label{xcm},
\end{equation}
\begin{equation}
\phi_0=c_3-{1\over2}{\dot\gamma}(t)^2\,x_{\rm
cm}(t)^2+{1\over2}\left({1\over4}({a\,N})^2+e^{-2\gamma(0)}\left(v_0+
x_0\,{\dot\gamma}(0)\right)^2\right)g(t),
\end{equation}
$g(t)=\int_{0}^t e^{2\gamma(t^\prime)}dt^\prime$. The constant $c_3$
corresponds to an arbitrary overall phase.

This solution corresponds to a sech-shaped soliton containing $N$
atoms with a time-dependent center-of-mass $x_{\rm cm}(t)$ and
time-dependent width $2\exp(-\gamma\,t)/N\,a$. The linear part of
the phase profile shows that the soliton is moving with a
center-of-mass velocity $v(t)={\dot x}_{\rm cm}(t)$. The quadratic
part corresponds to a phase chirp associated with the quadratic
trapping potential.

The simple choice $\gamma(t)=constant$, which corresponds to the
homogeneous Gross-Pitaevskii equation, gives $x_{\rm
cm}(t)=v_0\,t+x_0$. The solitonic solution in this case corresponds
to the well-known sech-shaped soliton with a center-of-mass moving
with a constant velocity $v_0$ and starting the motion at the
initial position $x_0$. In the limit $\gamma(t)\rightarrow 0$, the
solution, Eq.~(\ref{psi_exact}), reduces to
\begin{equation}
\psi(x,t)={\sqrt{N}\sqrt{N\,a}\over2}\,e^{i\,\phi(t)}{\rm
sech}{\left({N\,a\over2}(x-(x_0+v_0\,t))\right)}
\label{psi_exactg0},
\end{equation}
where
$\phi(t)=c_3+((a\,N)^2+4v_0^2)\,t/8+v_0\left(x-(x_0+v_0\,t)\right)$.
This is the well-known sech solution of the homogeneous
Gross-Pitaevskii equation. For $\gamma(t)=constant\times t$ we get a
Gross-Pitaevskii equation with an expulsive harmonic potential and
exponentially growing interatomic interaction \cite{liang}. In
principle, we can choose any form of $\gamma(t)$, but one should
keep in mind that the interatomic interaction strength is
proportional to $\exp{(\gamma(t))}$. Such a time-dependence may not
be realistic from an experimental point of view for any $\gamma(t)$.

In Ref.~\cite{serk} Serkin {\it et al.} have already derived the
exact solitonic solution of a Gross-Pitaevskii equation with a
linear and quadratic potentials and time-dependent dispersion
\begin{equation}
i{\partial\over\partial
t}\psi(x,t)=\left[-{D(t)\over2}{\partial^2\over\partial
x^2}-{\Omega^2(t)\over2}x^2-2\alpha(t)x-a\,
R(t)|\psi(x,t)|^2\right]\psi(x,t) \label{gp_serk}.
\end{equation}
The authors show that this equation is integrable only if the
functions $D(t)$, $\Omega(t)$ and $R(t)$  are related to each other
through the {\it integrability condition} (Eq.~(2) in
Ref.~\cite{serk}). The special case of only a quadratic potential
and constant dispersion corresponds to the Gross-Pitaevskii equation
considered in this paper. Substituting $\alpha(t)=0$ and $D(t)=1$ in
Eq.~(2) of Ref.~\cite{serk}, the integrability condition simplifies
to $\Omega^2=-{\dot\gamma}^2+{\ddot\gamma}$, which shows that, in
this special case, the previous equation is indeed equivalent to
Eq.~(\ref{gp2}). Therefore, substituting $\alpha(t)=0$ and $D(t)=1$
in the general solution of Eq.~(\ref{gp_serk}), namely Eq.~(8) in
Ref.~\cite{serk}, should result in our solution,
Eq.~(\ref{psi_exact}). It turns out, however, that the two solutions
do not match exactly. The solution of Ref.~\cite{serk} corresponds
to a soliton located initially at $x=0$  while in our case the
soliton is located initially at the arbitrary position $x_0$. This
can be clearly seen by substituting, without loss of generality,
$\eta(t)=N\,a\,\exp(\gamma(t))/4$ and
$\kappa(t)=-v_0\,\exp(-\gamma(0))/2$ in the argument of the sech
function of Eq.~(8) of Ref.~\cite{serk}. This results in a center of
mass coordinate $x_{\rm cm}(t)=v_0 g(t)\,e^{-\gamma(t)-\gamma(0)}$.
Since $g(0)=0$, this shows that $x_{\rm cm}(0)=0$ in contrast with
our case $x_{\rm cm}(0)=x_0$.

The dynamics of the soliton is readily given by Eq.~(\ref{xcm}). An
equation of motion for the center-of-mass $x_{\rm cm}(t)$ can be
derived from the Euler's equation of the lagrangian $L[x_{\rm
cm},{\dot x}_{\rm
cm}]=\int_{-\infty}^{\infty}i\,dx\,\psi^*{\partial\psi/\partial
t}-E[x_{\rm cm},{\dot x}_{\rm cm}] $. The energy functional is given
by
\begin{equation}
E[x_{\rm cm},{\dot x}_{\rm cm}]=\int_{-\infty}^{\infty}
\psi^*(x,t)\left[-{1\over2}{\partial^2\over\partial
x^2}+{1\over2}\,({\ddot \gamma}(t)-{\dot\gamma}(t)^2)\,
x^2-{1\over2}a\,e^{\gamma(t)}|\psi(x,t)|^2\right]\psi(x,t)
\label{efunc},
\end{equation}
which results in the equation of motion
\begin{equation}
{\ddot x}_{\rm cm}(t)+({\ddot\gamma}-{\dot\gamma}^2)x_{\rm cm}(t)=0
\label{xcm_eq}.
\end{equation}
This equation shows that the center-of-mass motion is determined by
the function $\gamma(t)$. Thus, interatomic interactions do not
affect the center-of-mass motion which is a manifestation of Kohn's
theorem \cite{kohn}.

\subsection{Oscillating trapping potential}
\label{subsecosc} 
\begin{figure}
\begin{center}
\includegraphics[width=15cm]{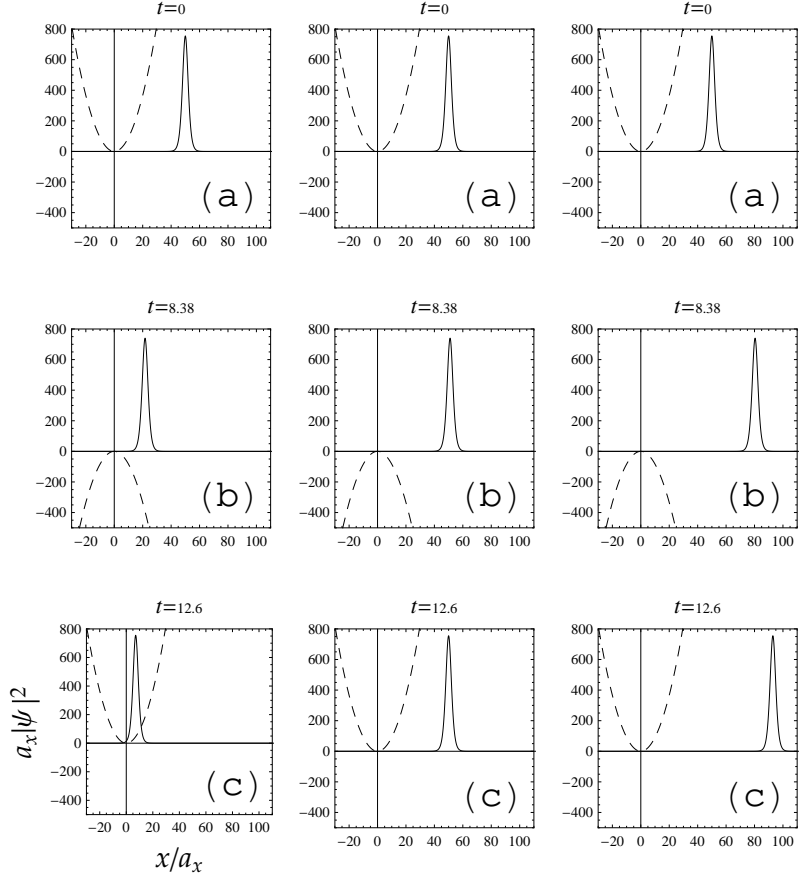}
\end{center}
\caption{Soliton's density profile (solid curve) and trapping
potential (dashed curve). First column of subfigures: $v_0=0.8\times
x_0\,\alpha_2\omega\,\sin{(\delta)}/2=8.4$. Middle column of
subfigures: $v_0=x_0\,\alpha_2\omega\,\sin{(\delta)}/2=10.5$. Last
column of subfigures: $v_0=1.2\times
x_0\,\alpha_2\omega\,\sin{(\delta)}/2=12.6$. For the three cases, we
take $\alpha_1=1$, $\alpha_2=0.5$, $\omega=2$, $\delta=1$, and
$x_0=50$. Time is in units of $1/\omega_x$.} \label{fig1}
\end{figure}

\begin{figure}
\begin{center}
\includegraphics[width=10cm]{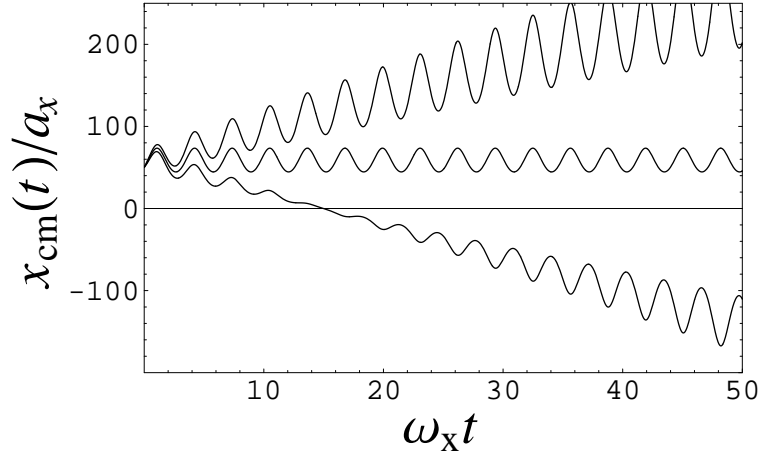}
\end{center}
\caption{Soliton's center-of-mass trajectory for three values of the
initial velocity. The upper curve corresponds to $v_0=1.2\times
x_0\,\alpha_2\omega\,\sin{(\delta)}/2=12.6$, the middle curve
corresponds to $v_0=x_0\,\alpha_2\omega\,\sin{(\delta)}/2=10.5$, and
the lower curve corresponds to $v_0=0.8\times
x_0\,\alpha_2\omega\,\sin{(\delta)}/2=8.4$. For the three curves, we
take $\alpha_1=1$, $\alpha_2=0.5$, $\omega=2$, $\delta=1$, and
$x_0=50$.} \label{fig2}
\end{figure}

In this section, we consider an oscillating form
of $\gamma(t)$ which results in a trapping potential and interatomic
interaction with oscillating strengths. A first simple choice for
$\gamma(t)$ would be for instance $\cos{(\omega\,t+\delta)}$.
However, in this case, the interatomic interaction, which will be
proportional to $\exp{(\cos{\omega\,t+\delta})}$, oscillates
nonlinearly. Such time-dependent interatomic interaction may not be
possible to realize experimentally. Instead, we use the form
\begin{equation}
\gamma(t)={1\over2}(\alpha_1+\alpha_2\cos{(\omega
t+\delta)})\label{gamma0},
\end{equation}
where $\alpha_1$, $\alpha_2$, $\omega$, and $\delta$ are arbitrary
dimensionless constants. In this case, the Gross-Pitaevskii
equation, Eq.~(\ref{gp2}), takes the form
\begin{equation}
i{\partial\psi(x,t)\over\partial
t}=\left[-{1\over2}{\partial^2\over\partial
x^2}-{1\over4}\alpha_2\omega^2 \left(\cos{(\omega
t+\delta)}+{1\over2}\alpha_2\sin{(\omega
t+\delta)^2}\right)x^2-a\,e^{(\alpha_1+\alpha_2\cos{(\omega
t+\delta)})/2}|\psi(x,t)|^2\right]\psi(x,t) \label{gp3}.
\end{equation}
The advantage of this particular form of $\gamma(t)$ is that, for
$\alpha_2\ll1$, the amplitude of the oscillation in the interatomic
interaction can be set to an arbitrarily small value such that the
strength of the interatomic interaction can be considered
practically as constant. Substituting this expression for
$\gamma(t)$ in Eq.(\ref{xcm_eq}), we get
\begin{equation}
{\ddot x}_{\rm cm}-{1\over4}\alpha_2\,\omega^2\,
\left(2\cos{(\omega\,t+\delta)}+\alpha_2\sin{(\omega\,t+\delta)}^2\right)x_{\rm
cm}=0 \label{xcm_eq20}.
\end{equation}
The general solution of this equation is readily given by
Eq.~(\ref{xcm}), which now takes the form
\begin{equation}
x_{\rm
cm}(t)=x_0\,e^{{\alpha_2\over2}(\cos{\delta}\,-\cos{(\omega\,t+\delta)})}
+\left(v_0-{1\over2}x_0\,\alpha_2\,\omega\,\sin{\delta}\right)e^{-{\alpha_2\over2}
(\cos{\delta}+\cos{(\omega\,t+\delta)})}\int_0^t{dt^\prime\,e^{\alpha_2\,\cos{(\omega\,t^\prime+\delta)}}}.
\label{xcm3}
\end{equation}
The first term of this equation corresponds to a bounded
oscillation, but the presence of the integral in the second term
makes $x_{\rm cm}(t)$ unbounded. Due to this term, the soliton will
be expelled out of the trap, i.e., $x_{\rm
cm}(t\rightarrow\infty)=\pm\infty$. Therefore, the soliton can be
localized, by choosing the parameters such that the prefactor of the
unbounded term vanishes, namely
\begin{equation}
v_0={1\over2}x_0\,\alpha_2\,\omega\,\sin{\delta}.
\end{equation}
With this condition, the center-of-mass of the soliton is given by
\begin{equation}
x_{\rm
cm}(t)=x_0\,e^{{\alpha_2\over2}(\cos{\delta}\,-\cos{(\omega\,t+\delta)})}
\label{xcm4}.
\end{equation}
This is one of the main conclusions of this paper. It shows that the
soliton can be localized at an arbitrary position by oscillating the
strength of the trapping potential and the interatomic interaction.
Without such oscillations, the soliton will be expelled out of the
trap. In fact, this result also holds for any bounded type of
oscillations as can be inferred from Eq.~(\ref{xcm}). Taking
$v_0+x_0\,{\dot\gamma}(0)=0$, this equation gives $x_{\rm
cm}(t)=x_0\,e^{\gamma(0)-\gamma(t)}$, which is bounded for any
bounded $\gamma(t)$. The different cases of soliton localization and
delocalization, described by Eq.~(\ref{xcm3}), are shown in
Figs.~\ref{fig1} and \ref{fig2}. In Fig.~\ref{fig1}, the soliton is
shown to be expelled out of the left (right) side of the trap when
$v_0<x_0\,\alpha_2\,\omega\,\sin{(\delta)}/2$
($v_0>x_0\,\alpha_2\,\omega\,\sin{(\delta)}/2$), while for
$v_0=x_0\,\alpha_2\,\omega\,\sin{(\delta)}/2$, the soliton remains
localized around its initial position $x_0$. In Fig.~\ref{fig2},
this is shown with the trajectory of the center-of-mass of the
soliton.

Notice that in order to localize the soliton, no condition on
$\omega$ was required. Therefore, one may argue that by taking
$\omega$ arbitrarily small, we get a localized soliton in an almost
stationary expulsive harmonic trap. This of course contradicts the
fact that in an expulsive harmonic trap, solitons are expelled away
from the center. However, one should keep in mind that with our
special form of $\gamma(t)$, namely Eq.~(\ref{gamma0}), the strength
of the trapping potential will be proportional to $\omega^2$.
Therefore, a very small value of $\omega$ corresponds to a shallow
potential that approaches the homogeneous case for $\omega=0$. The
fact that the strength of the harmonic potential depends on $\omega$
leads to conclude that, for deep traps, larger trap oscillations
frequency are needed to localize the soliton contrary to the case
with shallow traps where the soliton can be localized with smaller
trap oscillations frequency.

\begin{figure}
\begin{center}
\includegraphics[width=10cm]{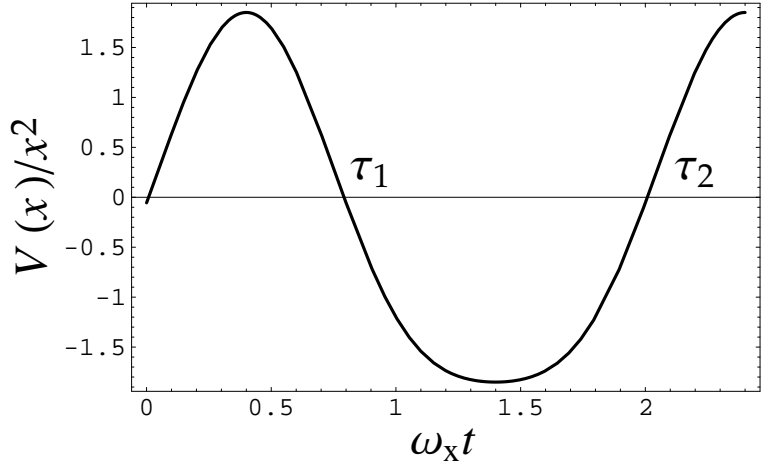}
\end{center}
\caption{ Strength of the trapping potential versus time.  The
following values were used: $\alpha_1=1$, $\alpha_2=0.75$,
$\omega=\pi$, $\delta=1.2$.} \label{fig3}
\end{figure}

Taking the time average of the trapping potential
$(\omega/2\pi)\int_0^{2\pi/\omega}d\,t\,({\ddot\gamma}(t)-{\dot\gamma}(t)^2)\,x^2/2
=-\alpha_2^2\,\omega^2\,x^2/16$ shows that the soliton spends on the
average more time in the expulsive trapping potential. One may thus
conclude that after sufficiently long time the soliton will be
expelled out of the trap which contradicts our previous localization
result. A careful examination of the dynamics shows that this
conclusion is incorrect. In spite of the fact that the soliton
spends more time in the expulsive trap, the inhomogeneity of the
trapping potential can compensate for the time difference. The
strength of the trapping potential has in general two periods
$\tau_1$ and $\tau_2$, as shown in Fig.~\ref{fig3}. The period
$\tau_1$ depends on $\alpha_2$ and $\tau_2=2\pi/\omega$. The
strength of the trapping potential is positive for a period of
$\tau_1$ and negative for $\tau_2-\tau_1>\tau_1$ for all $\alpha_2$.
Assuming the soliton started the motion at $x_0>0$ from rest, it
will experience at first a harmonic trapping potential for time
$\tau_1$ and therefore will move to the left (region of lower
potential) for a certain distance. At time $\tau_1$, the trapping
potential becomes expulsive for time period $\tau_2-\tau_1$ and the
soliton starts to return back. However, it starts now from a point
of lower potential than at $x_0$ which means that it will experience
a weaker trapping potential. Therefore, in order to reach the
starting point, $x_0$, it needs more time compared to the forward
part of the motion. If that time matches $\tau_2-\tau_1$, the
soliton returns back to $x_0$ with zero velocity and the cycle
repeats leading to trapping of the soliton, which corresponds to the
middle curve of Fig.~\ref{fig2}. On the other hand, if the soliton
reaches a point $x>x_0$, it will be eventually expelled out of the
right side of the trap corresponding to the upper curve of
Fig.~\ref{fig2}, and if it reaches $x<x_0$, it will be expelled out
of the left side of the trap corresponding to the lower curve of
Fig.~\ref{fig2}.

\begin{figure}
\begin{center}
\includegraphics[width=15cm]{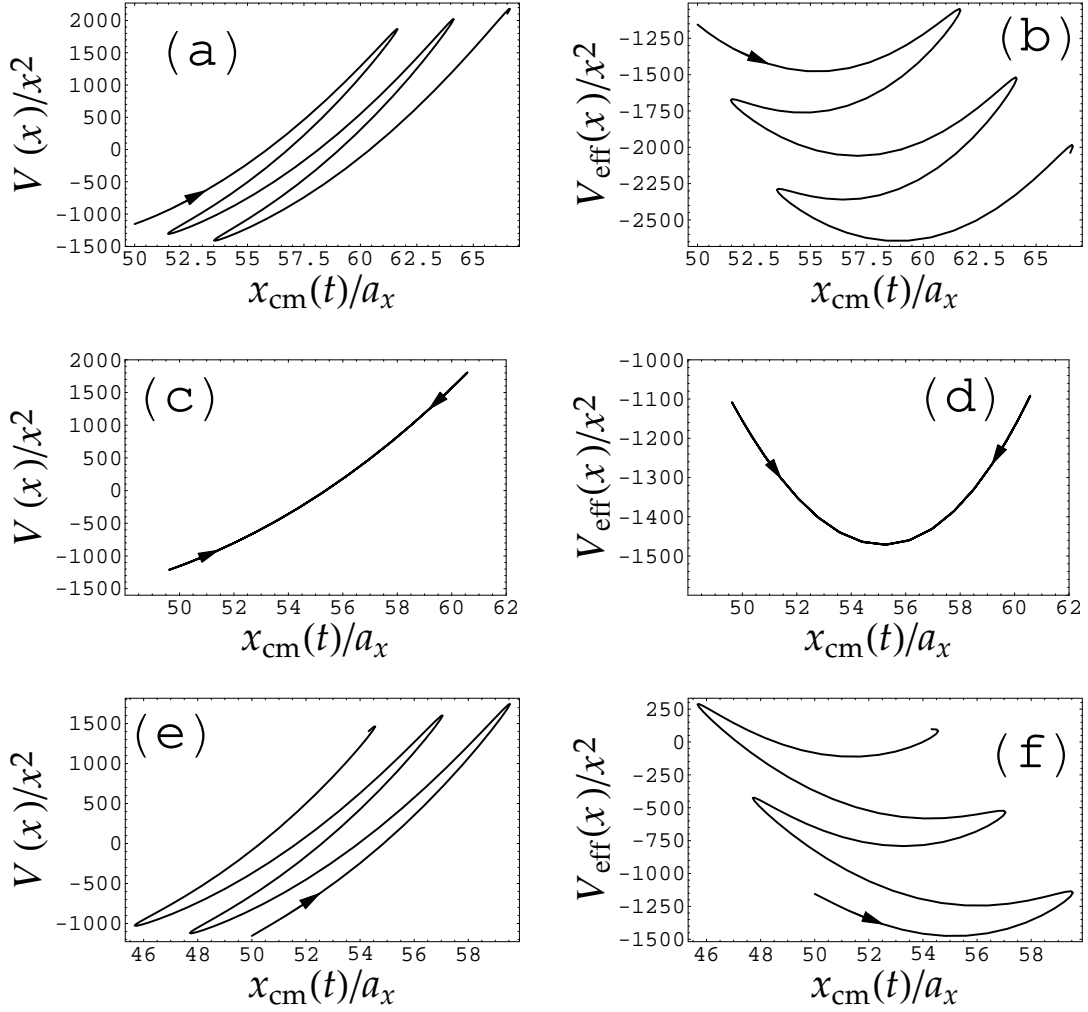}
\end{center}
\caption{First column: The trapping potential $V(x_{\rm cm}(t))$.
Second column: The effective trapping potential $V(x_{\rm cm}(t))$.
(a) and (b) correspond to,
$v_0=1.2\,\alpha_2\,x_0\,\omega\,\sin{(\delta)}/2$, (c) and (d)
correspond to, $v_0=\alpha_2\,x_0\,\omega\,\sin{(\delta)}/2$, (e)
and (f) correspond to,
$v_0=0.8\alpha_2\,x_0\,\omega\,\sin{(\delta)}/2$. For all curves:
$x_0=50$, $\alpha_1=1$, $\alpha_2=0.2$, $\delta=0.4$, and
$\omega=\pi$. The arrows show the direction of the motion of the
soliton.} \label{fig4}
\end{figure}

To further understand this trapping mechanism, we calculated the
trapping potential {\it felt} by the soliton through its trajectory,
namely $V(x_{\rm cm}(t))$, which is plotted in Fig.~\ref{fig4}. In
this figure, the dynamics of the soliton is represented by a point
moving on the curve with a direction that is indicated on each
curve. The center-of-mass motion of Fig.~\ref{fig2} can be extracted
by tracing $x_{\rm cm}(t)$  while the point moves along the
potential curves. In Fig.\ref{fig4}a, the soliton starts at $x_0=50$
with initial velocity
$v_0=1.2\,\alpha_2\,x_0\,\omega\,\sin{(\delta)}/2$. In this case,
the soliton gets drifted by time towards larger values of $x_{\rm
cm}$ corresponding to the upper curve of Fig.~\ref{fig1}. In
Fig.~\ref{fig4}c, the initial velocity is
$v_0=\,\alpha_2\,x_0\,\omega\,\sin{(\delta)}/2$, which results in
soliton localization. In this subfigure, the soliton oscillates
between $x_{\rm cm}=50$ and $x_{\rm cm}=61$ corresponding to the
middle curve of Fig.~\ref{fig2}. In Fig.~\ref{fig4}e, the initial
velocity is $v_0=0.8\,\alpha_2\,x_0\,\omega\,\sin{(\delta)}/2$
leading to a drift towards lower values of $x_{\rm cm}$
corresponding to the lower curve of Fig.~\ref{fig2}. To make an
analogy with a classical particle moving in a time-independent
potential, we defined $V_{\rm eff}(x_{\rm cm}(t))=V(x_{\rm
cm}(t))-[V(x_{\rm cm}(\pi/\omega))-V(x_{\rm cm}(0))][x_{\rm
cm}(t)-x_{\rm cm}(0)]/[x_{\rm cm}(\pi/\omega)-x_{\rm cm}(0)]$. In
Figs.~\ref{fig4}b, d, f, we plot $V_{\rm eff}(x_{\rm cm})$ that
corresponds to Figs.~\ref{fig4}a, c, e, respectively. The dynamics
is now simplified to that of a classical particle oscillating in a
{\it ladder} of parabolic potentials. For the nonlocalized soliton,
the potential minimum is shifted by time to the right
(Fig.~\ref{fig4}b) or to the left (Fig.~\ref{fig4}f). For the
localized soliton case, the minimum of the potential is stationary.

\subsection{Another family of exact solutions}
\label{subsecanother} 
\begin{figure}
\begin{center}
\includegraphics[width=10cm]{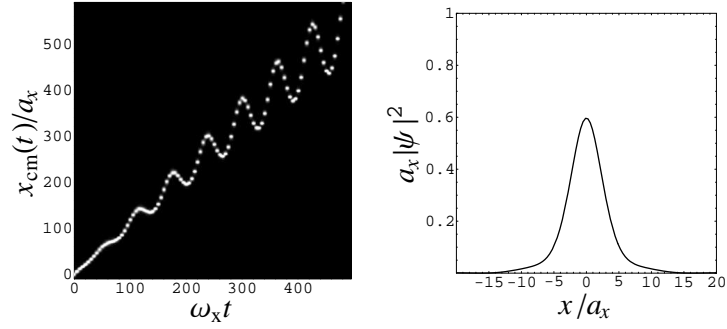}
\end{center}
\caption{(Left) Spatiotemporal contour plot of soliton density
profile. (Right) Soliton density profile at $t=0$. The values of the
parameters used in this plot are: $c_1=-c_2=10$, $c_3=c_4=1$,
$\lambda_{1i}=\lambda_{1r}=1$, $A=2$, $a=0.9$, $k_0=5$,
$\lambda=\omega=1$, $\delta=0$, $\alpha_1=-6$, $\alpha_2=0.3$. The
value of $k_0$ is the solution of $\alpha=0$ with respect to $k_0$.}
\label{fig5}
\end{figure}

\begin{figure}
\begin{center}
\includegraphics[width=10cm]{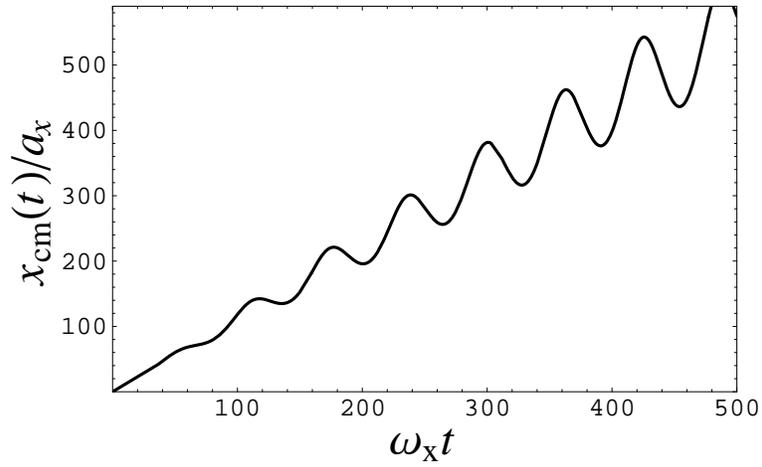}
\end{center}
\caption{Trajectory of the soliton's peak density. The values of the
parameters used here are the same as those of Fig.5.} \label{fig6}
\end{figure}

\begin{figure}
\begin{center}
\includegraphics[width=10cm]{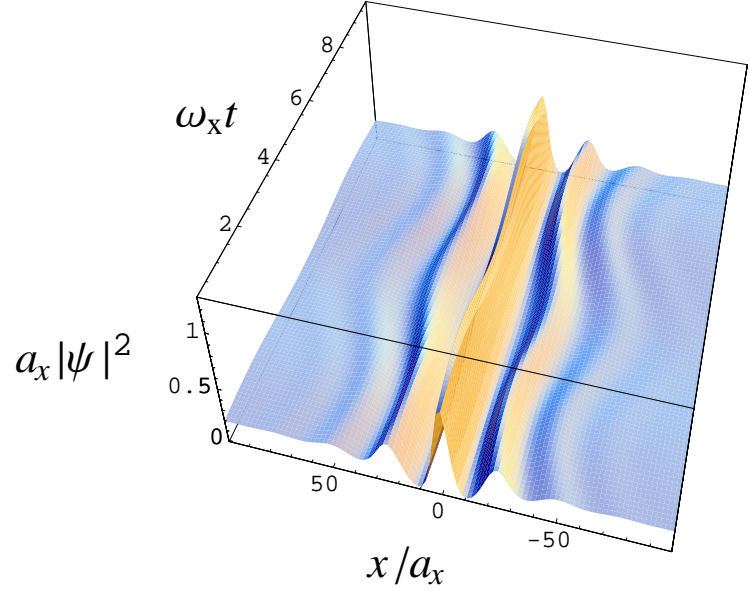}
\end{center}
\caption{(Color online) Density profile of a single soliton
solution. The values of the parameters used in this plot are:
$-c_2=c_1=10$, $c_3=c_4=1$, $\lambda_{1i}=\lambda_{1r}=1$, $A=2$,
$a=0.9$, $k_0\sim5.03$, $\lambda=\omega=1$, $\delta=0$,
$\alpha_1=-6$, $\alpha_2=0.3$. The value of $k_0$ is the solution of
$\alpha=0$ with respect to $k_0$.} \label{fig7}
\end{figure}

\begin{figure}
\begin{center}
\includegraphics[width=10cm]{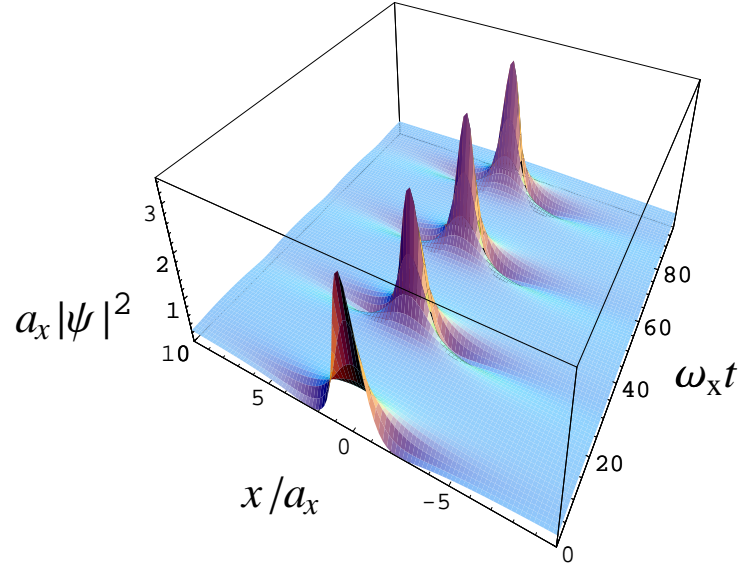}
\end{center}
\caption{(Color online) Density profile of a multi-solitons
solution. The values of the parameters used in this plot are:
$-c_2=c_1=10$, $c_3=c_4=1$, $\lambda_{1i}=0$, $\lambda_{1r}=1$,
$A=2$, $a=0.9$, $k_0=0$, $\lambda=1$, $\omega=0.01$, $\delta=0$,
$\alpha_1=-2$, $\alpha_2=0.3$.} \label{fig8}
\end{figure}

Using the exact solution found above as a seed
solution, the Darboux transformation generates a two-solitons
solution. This kind of solution is useful for studying
soliton-soliton interactions which is left for future work. Another
family of more complicated exact solitonic solutions can be obtained
by using a nontrivial seed solution as shown in Appendix A.
Substituting for $\gamma(t)$ in Eq.~(\ref{gpsol}), we get the second
class of exact solutions. This family of solutions is more
complicated than the above single-solitonic solution since it
involves, in addition to the single-solitonic solutions,
multi-solitonic and solitary wave solutions. Here, we present
briefly the main properties of the solutions of this kind. In
Fig.~\ref{fig5}, we plot the density of a single-solitonic solution
showing that the soliton is being expelled out of the center of the
trap. The oscillation in the trajectory of the soliton is due to the
oscillating trapping potential. The discontinuous appearing of the
soliton is due to the interaction with the background. This
trajectory can also be extracted from the general solution,
Eq.~(\ref{gpsol}), by considering the term $(\Gamma
c_1^2e^{2\alpha\eta-\sqrt{\dot\eta}\Delta_rx}+\Gamma
c_2^2e^{-2\alpha\eta+\sqrt{\dot\eta}\Delta_rx})$ in the denominator.
At the soliton's density peak this is the dominant term that
determines the position of the soliton. Specifically, the position
of the peak is given by the condition
$2\alpha\eta-\sqrt{\dot\eta}\Delta_rx=0$. Using this condition to
plot $x$ versus $t$ in Fig.~\ref{fig6}, we obtain a curve that is
identical to the soliton trajectory in Fig.~\ref{fig5}. The mean
slope of this curve is proportional to $\alpha/2\Delta_r$. Hence,
the rate at which the soliton leaves the center of the trapping
potential can be delayed by choosing the parameters and the
arbitrary constants such that $\alpha/2\Delta_r$ is small. For the
special case of $\alpha=0$ the center-of-mass of the soliton will be
localized at $x=0$ indefinitely. In this case, the oscillating
trapping potential results only in oscillations in the width and
peak density of the soliton. This is also shown in Fig.~\ref{fig7}
where we see a multi-solitonic solution with central soliton being
localized at $x=0$ and off-central ones oscillating around their
initial positions. In Fig.~\ref{fig8}, we show that for some values
of the parameters the dynamics of the peak soliton density can be so
drastic such that the soliton disappears in the background and
reappears at regular discrete times.

\section{Numerical solution and experimental realization}
\label{secnum} 
\begin{figure}
\begin{center}
\includegraphics[width=10cm]{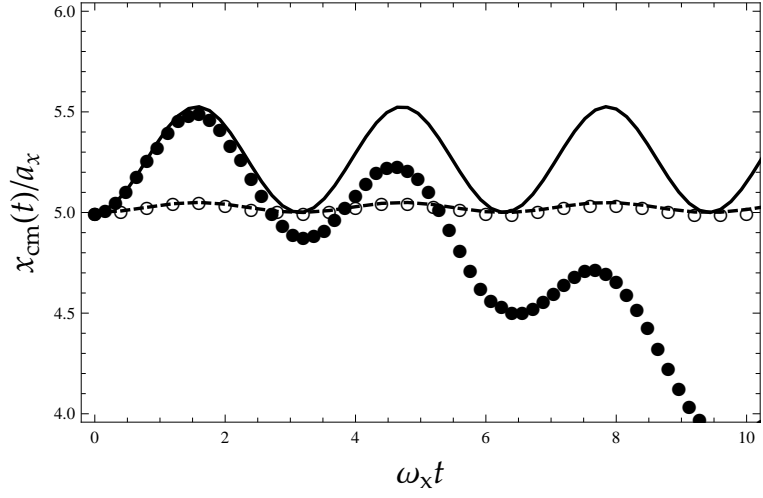}
\end{center}
\caption{Soliton's center-of-mass trajectory. The solid and dashed
curves correspond to the exact formula Eq.~(\ref{xcm}). Empty and
filled circles correspond to the numerical solution of
Eq.~(\ref{gp4}). Solid curve and filled circles are obtained with
$\Omega=\omega\sqrt{\alpha_2/2}=0.44\,\omega_x\,$ and dashed curve
and empty circles are obtained with
$\Omega=\omega\sqrt{\alpha_2/2}=0.14\,\omega_x$. The rest of
parameters used are: $a=10^{-4}$, $\delta=v_0=0$, $x_0=5$,
$\alpha_1=1$, $N=4\times10^{3}$, $\alpha_2=0.1$ for the solid curve
and $\alpha_2=0.01$ for the dashed curve.} \label{fig9}
\end{figure}

As we have seen in section~\ref{subsecosc}, there is
a possibility to localize the soliton by oscillating the trapping
potential and the interatomic interaction in the manner described by
Eq.~(\ref{gp3}). Such synchronized oscillations may not be possible
to realize experimentally. Instead, a setup with an oscillating
strength of the trapping potential and constant interatomic
interaction may be experimentally more favorable. This situation can
be obtained in our case with the condition $\alpha_2\ll1$ resulting
in the following Gross-Pitaevskii equation
\begin{equation}
i{\partial\psi(x,t)\over\partial t}=\left[-{\partial^2\over\partial
x^2} -{1\over2}\Omega^2 \cos{(\omega
t+\delta)}\,x^2-a\,|\psi(x,t)|^2\right]\psi(x,t) \label{gp4}
\end{equation}
where $\Omega=\omega\,\sqrt{\alpha_2/2}$ and we have set
$\alpha_1=0$. We solve this equation numerically using the exact
solution of the homogeneous case, namely the solution of
Eq.~(\ref{gp4}) with $\Omega=0$, as the initial wavefunction. This
can be obtained from Eq.~(\ref{psi_exact}) simply by substituting
$\alpha_1=\alpha_2=0$. The soliton's center-of-mass trajectory is
extracted from the resulting numerical solution and then plotted
versus time as shown in Fig.~\ref{fig9}. The trajectory is shown
with the filled circles for $\omega=2$ and $\Omega=0.44$ and with
the empty circles for $\omega=2$ and $\Omega=0.14$. It is clear from
this figure that with smaller amplitude of the oscillating trapping
potential, the soliton will be localized for longer periods. The
solid and dashed curves show the corresponding trajectories in the
presence of the oscillating interatomic interaction as described by
Eq.~(\ref{xcm3}). The difference between the solid and filled
circles curves shows the important role played by the oscillations
in the interatomic interactions in stabilizing the soliton. On the
other hand, the overlap between the dashed and open circles curves
shows that the effect of interatomic interactions is minor for
smaller amplitudes of the oscillation in the trapping potential.

\begin{figure}
\begin{center}
\includegraphics[width=10cm]{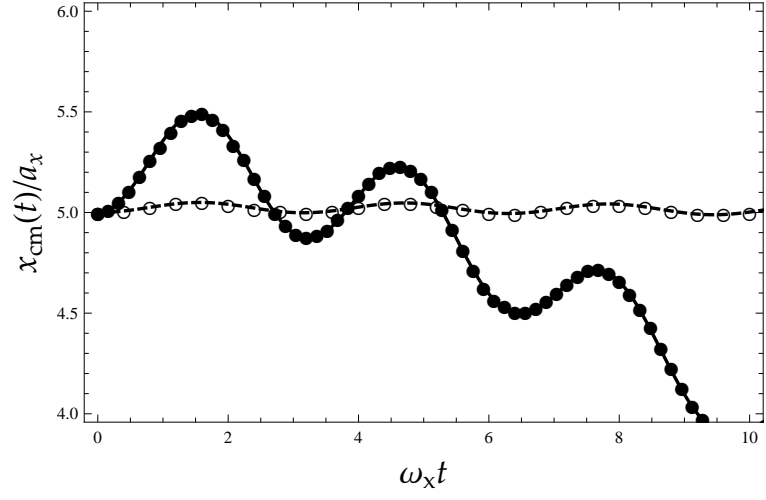}
\end{center}
\caption{Soliton's center-of-mass trajectory. The solid and dashed
curves correspond to the exact formula Eq.~(\ref{math}). Empty and
filled circles correspond to the numerical solution of
Eq.~(\ref{gp4}). All parameters used are are the same as those of
Fig.~\ref{fig9}.} \label{fig10}
\end{figure}

\begin{figure}
\begin{center}
\includegraphics[width=10cm]{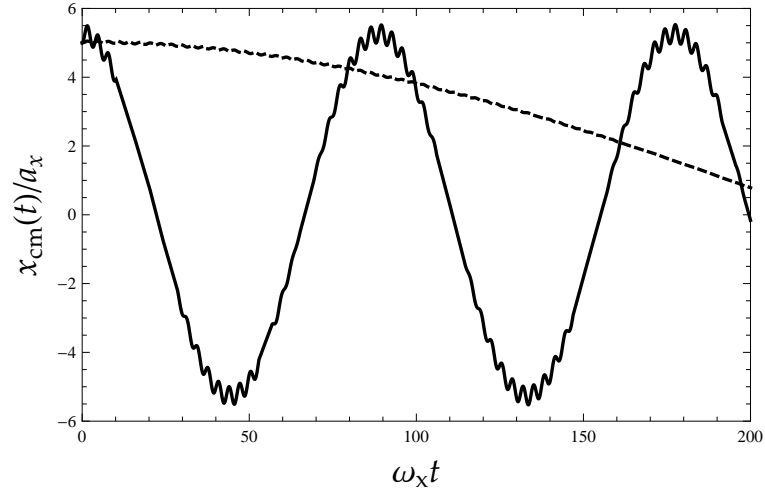}
\end{center}
\caption{The solid and dashed curves correspond to the exact formula
Eq.~(\ref{math}). The two curves are the same as those of
Fig.~\ref{fig10} but shown here over a larger time interval.
Extending the time interval further shows that the dashed curve is
also oscillatory.} \label{fig11}
\end{figure}

The exact center-of-mass dynamics of the solitonic solution of
Eq.~(\ref{gp4}) is dictated, according to Kohn's theorem
\cite{kohn}, by the potential $-{1\over2}\Omega^2 \cos{(\omega
t+\delta)}\,x^2$ independently from the interatomic interaction.
Taking advantage of this fact, the equation of motion of the center
of mass follows
\begin{equation}
{\ddot x_{\rm cm}}(t)-\Omega^2\,\cos{(\omega\,t+\delta)}\,x_{\rm
cm}(t)=0.
\end{equation}
The general solution of this equation is a linear combination of the
sine and cosine Mathieu functions
\begin{equation}
x_{\rm
cm}(t)=c_1\,C(0,\alpha_2,(\omega\,t+\delta)/2)+c_2\,S(0,\alpha_2,(\omega\,t+\delta)/2),
\end{equation}
where $c_1$ and $c_2$ are arbitrary constants. Using the initial
conditions $x_{\rm cm}(0)=x_0$ and ${\dot x}_{\rm cm}(0)=0$, the
solution takes the form
\begin{equation}
x_{\rm cm}(t)={x_0\,C(0,\alpha_2,(\omega\,t+\delta)/2)\over
C(0,\alpha_2,\delta/2)}\label{math}.
\end{equation}
This solution is plotted in Fig.~\ref{fig10}, where we also plot the
result of the numerical solution of Eq.~(\ref{gp4}) for the same
parameters. The agreement between the numerical and the exact result
is evident. The advantage of the exact analytical solution over the
numerical one is that we can investigate the long-time dynamics of
the soliton. In Fig.~\ref{fig11}, we plot the center-of-mass of the
soliton for a much longer time interval than in Figs.~\ref{fig9} and
\ref{fig10}. This figure shows that the soliton will be trapped over
such a large time scale and is oscillating between $x_0$ and $-x_0$.
The frequency of this oscillation is given by the period of the
Mathieu function $C(0,\alpha_2,(\omega\,t+\delta)/2)$. A numerical
computation of the first root of this function for different values
of $\alpha_2$ and $\omega$ shows that this frequency is proportional
to $\alpha_2\,\omega$. The constant of proportionality is determined
numerically which results in $\omega_P=0.353\,\alpha_2\,\omega$. The
fact that the soliton is trapped by the oscillating harmonic
potential is a well-established result for such a configuration,
known as {\it Paul trap} \cite{paul}, which is used to trap cold
ions. (Hence, $\omega_P$ denotes the frequency of the Paul trap.).

To have realistic estimates of the parameters $\alpha_1$,
$\alpha_2$, and $\omega$, we consider the experiment of Strecker
{\it et al.} \cite{randy}. In this experiment, solitons were created
with a maximum number of $N=5000$ $^7$Li atoms per soliton. The
solitons' center-of-mass oscillated with amplitude
$\sim370\,{\rm\mu}$m and period $~310$ ms. The strength of the
harmonic trapping potential in the radial direction
$\omega_\perp=2\pi\times800$ rad/s was much larger than that of the
axial direction $\omega_x=2\pi\times3$ rad/s. In this case, the unit
of length used in this paper is $a_x\approx2\,{\rm\mu}$m and the
unit of time is $1/\omega_x\approx50$ ms. Furthermore, for the
$^7$Li scattering length $a_s\sim3a_0=1.5\times10^{-10}$ m, our
scaled scattering length is $a=a_s/a_x\simeq10^{-4}$. In view of
these experimental values, Fig.~\ref{fig9} is explained as follows.
The filled circles show a soliton located initially at
$10\,{\rm\mu}$m from the trap center. Oscillating the trapping
potential with frequency
$\Omega=0.44\,\omega_x\simeq2\pi\times1.3\,$ rad/s, the soliton will
be drifted a distance of $10\,{\rm\mu}$m from its initial location
in a time period of 1.2 s. On the other hand, using a rather more
{\it gentle} oscillation in the trapping potential, namely with
$\Omega=0.14\,\omega_x\simeq\,2\pi\times0.4\,$ rad/s, the soliton
will be nearly localized about its initial position over the same
time period.

The exact result, Eq.~(\ref{xcm3}), indicates that the soliton can
be trapped at any position and with any trap frequency as long as
the trapping potential and the interatomic interaction are
oscillating coherently. Furthermore, it shows that the soliton
maintains its single-solitonic structure irrespective of the
robustness of the trap oscillation. In the present case, where the
interatomic interactions oscillation is turned off, the situation
may be different. The soliton looses the possibility of indefinite
localization and may also loose its single-soliton structure.
Solving Eq.~(\ref{gp4}) for larger values of $\Omega$ and $x_0$
shows that the soliton will leave its initial position faster and
gets fragmented into many solitons that collide and interfere with
each other. Therefore, localizing the soliton at larger distances
while maintaining its single-solitonic structure requires shallower
traps.

\section{conclusions}
\label{conc} Using the Darboux transformation method, we derived
exact solitonc solutions of a class of Gross-Pitaevskii equations
represented by Eq.~(\ref{gp2}). The solutions are obtained for a
general time-dependent strength of the harmonic trapping potential
and a related time-dependent strength of the interatomic
interaction. Two classes of exact solitonic solitions were found.
The first class represents a single soliton with an arbitrary phase,
initial position, and initial velocity, as given by
Eq.~(\ref{psi_exact}). The second class comprises single, multiple,
and solitary wave solitonic solutions.

As a specific case, we considered an oscillating trapping potential
and interatomic interaction, as given by Eqs.~(\ref{gamma0}) and
(\ref{gp3}). We found that the soliton can be localized at an
arbitrary position and for any amplitude and frequency of the
oscillating trapping potential. This localization is possible in
spite of the asymmetric oscillation where the trapping potential
spends more time being expulsive. In such a case, one expects that
by time the soliton will be expelled out of the trap. It turns out,
however, that the inhomogeneity in the trapping potential has a
balancing effect such that it becomes possible to localize the
soliton indefinitely. As a consequence of the previously-mentioned
approximate nature of the Gross-Pitaevskii equation, the fact that
our solutions have a $\delta$-function center-of-mass is also
approximate \cite{new100}. For finite number of atoms, the
center-of-mass spreads which may lead to delocalization of the
soliton. For small variations in the center-of-mass, the soliton is
expected to remain localized, but when the center-of-mass spreading
is larger than the amplitude of the soliton oscillation around its
equilibrium point, we expect that localization disappears
completely.

To discuss the experimental realization, we considered a simpler
situation with an oscillating trapping potential and constant
interatomic interaction, as described by Eq.~(\ref{gp4}). The
numerical solution of this equation showed that soliton localization
is possible but for a finite time that can be controlled by the
frequency and amplitude of the trapping potential oscillation. It is
shown that for the $^7$Li experiment of Strecker {\it et al.}
\cite{randy}, the soliton can be localized for a time long enough to
be observed. With smaller frequency and amplitude, the localization
time becomes even larger.

The Gross-Pitaevskii equation provides an accurate description of
the dynamics of solitons as long as finite-temperature effects are
suppressed and atom losses are negligible \cite{book}. At finite
temperatures and with atom losses, the soliton broadens and starts
to loose it particle-like behavior. When exposed to oscillations in
the trapping potential, the soliton will, in this case, be
fragmented and soliton localization may not hold.

Using the localization mechanism, solitons can be prepared in
arbitrary initial conditions. This may be useful for studying
soliton-soliton collisions or the interaction of solitons with
potentials. For instance, the shallow oscillating trap can be turned
on immediately after the solitons were created which leads to
localizing the soliton at a certain position. Then by switching on
and off the oscillating trap the soliton can be moved from one point
to the other.

\section*{Acknowledgement} The author would like to acknowledge Vladimir N. Serkin
for useful discussions and helpful suggestions.

\appendix
\section{Exact Solutions Using a Nonzero Seed}
A nontrivial seed solution can be easily obtained by substituting in
Eq.~(\ref{gp2}) $\psi(x,t)=\exp{(h_1(t)+ih_2(x,t))}$, where $h_1(t)$
and $h_2(x,t)$ are real functions:
\begin{equation}
\psi_0(x,t)=A\,\exp{\left[\frac{ \,{\dot\gamma}(t)}{2} - {\frac{i
}{4}\,
        \left( 4\,a\,A^2 - 4\,e^{ \,{\dot\gamma}(t)}\,x \,{k_0} - 2\,{{k_0}}^2 +
          4 \,\int e^{2 \,{\dot\gamma}(t)}\,dt\,\left( -2\,a\,A^2 + {{k_0}}^2 \right)  + x^2\,{\ddot\gamma}(t) \right)
        }\right]}
\label{seed}.
\end{equation}
Here $A$ and $k_0$ are arbitrary constants.

Solving the linear system (\ref{psi_x}) and (\ref{psi_t}) using the
seed solution, Eq.~(\ref{seed}), and then substituting for
$\psi_0(x,t)$, $\psi_1(x,t)$, and $\phi_1(x,t)$ in the last
equation, we obtain the following new exact solution of
Eq.~(\ref{gp2}):
\begin{eqnarray}
\psi(x,t)&=&{\dot\eta}^{1/4}e^{-i{\ddot\eta}x^2/8{\dot\eta}}
\left\{\frac{}{}
Ae^{q_1}\right.\nonumber\\&+&4\lambda_{1r}e^{i\theta_4+q_2}
(2ic_3A\sqrt{a}e^{2\alpha\eta}+c_4q_3e^{-2i\theta_1+\Delta_r
\sqrt{\dot\eta}x})(c_3q_3^*e^{2\alpha\eta}-2iA\sqrt{a}
c_4e^{-i\theta_3+\Delta_r\sqrt{\dot\eta}x})\nonumber\\
&/&\left[c_1^2\Gamma e^{2\alpha\eta-\Delta_r\sqrt{\dot\eta}x}+c_2^2\Gamma
e^{-2\alpha\eta+\Delta_r\sqrt{\dot\eta}x}-4Ac_1c_2\sqrt{a}[(2\lambda_{1r}-\Delta_r)(\cos{2\theta_1}
+\cos{\theta_3})\right.\nonumber\\&+&\left.
(\Delta_i-2\lambda_{1r})(\sin{2\theta_1}+\sin{\theta_3})+(\sin{2\theta_1}+\sin{\theta_3})k_0)
\frac{}{}\right]\left.\frac{}{}\right\} \label{gpsol},
\end{eqnarray}
where\\
\begin{math}
q_1=i[2A^2a(2\eta-1)+k_0(k_0(1-2\eta)+2\sqrt{\dot\eta})x]/2
\end{math},\\
\begin{math}
q_2=-2\alpha\eta-\Delta\sqrt{\dot\eta}x
\end{math},\\
\begin{math}
q_3=\Delta_i+k_0+i(\Delta_r+2i\lambda_{1i}-2\lambda_{1r})
\end{math},\\
\begin{math}
\theta_1=\Delta_r\lambda_{1r}\eta-((2\lambda_{1i}+k_0)-\sqrt{\dot\eta})\Delta_i/2
\end{math},\\
\begin{math}
\theta_2=(k_0^2(1-2\eta)-4\Delta_r\lambda_{1r}\eta)/4
+2A^2a(1-2\eta)+2\Delta_i(2\lambda_{1i}\eta-\sqrt{\dot\eta}x)+2
k_0(\Delta_{i}\eta+\sqrt{\dot\eta}x)
\end{math},\\
\begin{math}
\theta_3=(2\Delta_i\lambda_{1i}-2\Delta_r\lambda_{1r}+\Delta_ik_0)\eta-\sqrt{\dot\eta}\Delta_ix
\end{math},\\
\begin{math}
\theta_4=k_0\sqrt{\dot\eta}x+(2A^2a-k_0^2)(2\eta-1)/2
\end{math},\\
\begin{math}
\Gamma=(\Delta_i-2\lambda_{1i})^2+(\Delta_r-2\lambda_{1r})^2+4A^2a
+k_0(2\Delta_i-4\lambda_{1i}+k_0)
\end{math},\\
\begin{math}
\alpha=\Delta_r\lambda_{1i}+\Delta_i\lambda_{1r}+\Delta_rk_0/2
\end{math},\\
\begin{math}
\Delta_r={\rm Re}[\sqrt{(2\lambda_1-ik_0)^2-4A^2a}]
\end{math},\\
\begin{math}
\Delta_i={\rm Im}[\sqrt{(2\lambda_1-ik_0)^2-4A^2a}]
\end{math},\\
\begin{math}
\gamma=\Delta_i+k_0+i(\Delta_r+2i\lambda_{1i}-2\lambda_{1r})
\end{math},\\
\begin{math}
\eta(t)=\int{e^{2{\dot\gamma}(t)}dt}
\end{math},\\
and $c_{1}$ and $c_2$ are arbitrary constants. It should be noted
that this is an exact solution of Eq.~(\ref{gp2}) for any
$\gamma(t)$.

There are 5 arbitrary parameters in the general solution, namely
$k_0$, $A$, $\lambda_1$, $c_1$, and $c_2$. The first three
parameters control the phase and amplitude of the seed solution
which is part of the general solution. The last two parameters
control the amplitude and phase of the general solution. The
solitonic solutions represented by Eq.~(\ref{gpsol}) are nonsingular
for all $x$ and $t$ since the denominator of this equation does not
vanish. This can be easily deduced from Eq.~(\ref{psinew}) where we
see that the denominator of Eq.~(\ref{gpsol}) is merely the
amplitude of $\phi_1$ and $\psi_1$ that vanishes only if we have the
trivial solution with $c_1=c_2=0$.

\newpage
\protect

\clearpage

\newpage
\protect



\begin{thebibliography}{99}
\bibitem{note} Since the new solution can already be extracted from
Eq.~(\ref{u0p}), there is no need to write the more complicated
equation for ${\bf V}[1]$ which leads to the same result.
\end{thebibliography}
\end{document}